\documentclass[showpacs,twocolumn,floats,superscriptaddress]{revtex4}
\usepackage{graphicx}
\usepackage{amsmath}
\usepackage{bm}
\usepackage{color}

\newcommand{\de}{\delta}

\newcommand{\ignore}[1]{\relax}
\begin{document}
\title{Microscopic Theory for the Quantum to Classical Crossover 
in Chaotic Transport}
\author{Robert S. Whitney}
\author{Ph.~Jacquod}
\affiliation{D\'epartement de Physique Th\'eorique,
Universit\'e de Gen\`eve, CH-1211 Gen\`eve 4, Switzerland}
\date{August 23, 2004}
\begin{abstract}
We present a semiclassical theory for the scattering matrix ${\cal S}$
of a chaotic ballistic cavity at finite Ehrenfest time. Using a phase-space 
representation coupled with a multi-bounce expansion, we show how the
Liouville conservation of phase-space volume decomposes ${\cal S}$ as 
${\cal S}={\cal S}^{\rm cl} \oplus {\cal S}^{\rm qm}$. The short-time, 
classical contribution ${\cal S}^{\rm cl}$ generates deterministic
transmission eigenvalues $T=0$ or $1$, while quantum ergodicity is recovered 
within the subspace corresponding to the long-time, stochastic contribution 
${\cal S}^{\rm qm}$. This provides a microscopic foundation for the two-phase 
fluid model, in which the cavity acts like a classical and a quantum cavity in 
parallel, and explains recent numerical data showing the breakdown of 
universality in quantum chaotic transport in the deep semiclassical limit.
We show that the Fano factor of the shot-noise power
vanishes in this limit, while weak localization remains universal.
\end{abstract}
\maketitle
Quantum mechanics strongly alters the classical theory of transport. In 
mesoscopic systems, spectacular new phenomena of purely quantum origin emerge, 
such as weak localization, universal conductance fluctuations \cite{Stone}, 
and sub-Poissonian shot-noise \cite{BlanterPR}. In both mesoscopic
disordered diffusive metals and clean chaotic ballistic systems,
these phenomena have been found to exhibit a universality well captured by 
Random Matrix Theory (RMT) \cite{carlormt}. For instance, in a 
time-reversal-symmetric chaotic cavity, RMT correctly predicts that weak 
localization reduces the conductance below its classical value by
$\delta g_{\rm rmt}=-1/4$, that the variance of the conductance
$\sigma_{\rm rmt}^2(g)=1/8$ (conductances are expressed
in units of $e^2/h$), and that the shot-noise power 
is reduced by the Fano factor $F_{\rm rmt}=1/4$, below the Poissonian value
of $2 e \langle I \rangle$, $\langle I \rangle$ being the average
current through the sample. 

Despite the many successes of RMT it has become clear, 
following the work of Aleiner and Larkin \cite{aleiner1},
that universality is broken for those quantum chaotic systems
in which the Ehrenfest time becomes a relevant time scale. 
The Ehrenfest time, $\tau_{\rm E}$, corresponds to the time it takes for a 
minimal wavepacket to be stretched over a length comparable to a classical 
characteristic length of the system. For a closed chaotic cavity of linear 
size $L$, the Ehrenfest time is
$\tau_{\rm E}^{\rm c} \simeq \lambda^{-1} \ln[L/\lambda_{\rm F}]$, 
in term of the classical cavity's Lyapunov exponent $\lambda$ and 
the Fermi wavelength $\lambda_{\rm F}$ \cite{Zaslavsky}.
For the same cavity attached to two-leads of width $W$ there is a second
shorter Ehrenfest time $\tau_{\rm E}^{\rm o} = \tau_{\rm E}^{\rm c} 
- 2\lambda^{-1} \ln[L/W]$ \cite{VavilovLarkin}.
These two quantum mechanical time scales differ only 
by a classical quantity. For times shorter than the Ehrenfest time,
the quantum mechanical time evolution of a narrow
wavepacket is given by the solution of classical equations
of motion \cite{Zaslavsky,steve}, thus strong deviations
from the RMT of transport emerge once $\tau_{\rm E}^{\rm o}$ is no longer
negligible. It is for example well established 
that, as $\tau_{\rm E}^{\rm o,c} \rightarrow \infty$, 
the Fano factor disappears \cite{agam,eugene2,jakub1,eugene} and
parametric conductance fluctuations remain universal 
while sample-to-sample conductance fluctuations do not
\cite{eugene,jakub2}. However the analytical prediction of 
Refs.~\cite{aleiner1,inanc} that weak localization corrections also
disappear has recently been challenged numerically \cite{jakub4}.

These observations are all consistent with the phenomenological
two-phase fluid model \cite{silvestrov,eugene}. In this model, introduced 
in Ref.~\cite{silvestrov}, it is assumed that electrons with short dwell 
time $< \tau_{\rm E}^{\rm o}$ are transmitted deterministically 
through the cavity, while quantum mechanical stochasticity is recovered for
those with longer dwell time $> \tau_{\rm E}^{\rm o}$. Accordingly, the
system splits into two cavities {\em in parallel}; 
a classical, deterministic cavity, with an effective
number of modes $N_{\rm cl}=N (1-\exp[-\tau_{\rm E}^{\rm o}/\tau_{\rm D}])$,
and a quantum mechanical, stochastic cavity having
$N_{\rm qm}=N \exp[-\tau_{\rm E}^{\rm o}/\tau_{\rm D}]$ modes 
($\tau_{\rm D}$ is the average dwell time). 
With the further assumption that the stochastic cavity obeys RMT
\cite{eugene,silvestrov}, the observed numerical behavior of shot-noise and 
of conductance fluctuations follows \cite{jakub1,eugene,jakub2}. The universal 
value of parametric conductance fluctuations is maintained as long as 
$N_{\rm qm} \propto \hbar_{\rm eff}^{-1+1/\lambda \tau_{\rm D}}\gg 1$,
where the effective Planck's constant $\hbar_{\rm eff} \equiv \lambda_F/L$
($\lambda_F$ is the Fermi wavelength).
For fully developed chaotic dynamics $\lambda \tau_{\rm D} \gg 1$,
and so this condition for $N_{\rm qm}$ 
is always satisfied in the semiclassical limit of 
$\hbar_{\rm eff} \rightarrow 0$ \cite{eugene}. 
However the prediction that weak localization remains universal (independently 
of $\hbar_{\rm eff}$) is in direct contradiction of the analytical prediction 
$\delta g \propto \exp[-\tau_{\rm E}^{\rm c}/\tau_{\rm D}]$ of 
Refs.~\cite{aleiner1,inanc}. Both this controversy and the successes of the 
phenomenological two-phase fluid model call for a microscopic foundation of 
this model. The purpose of this letter is
to provide such a foundation. 

We consider an open, two-dimensional chaotic quantum dot, ideally connected 
to two external leads, each carrying $N \gg 1$ modes. We require that the size 
of the openings to the leads is much smaller than the perimeter of the cavity 
but is still semiclassically large, $1 \ll N \ll L/\lambda_F$.
This ensures that the chaotic dynamics inside the dot has enough time to 
develop, i.e.~ $\lambda \tau_{\rm D} \gg 1$. The classical dynamics within 
the cavity can be described by its Birkhoff map, which provides a recursive
relation for the transverse momentum $p_{\tau}$ and position $q_{\tau}$ of 
incidence at the $\tau^{\rm th}$ collision with the cavity's boundary. 
Quantizing this map leads to a unitary, $M \times M$ Floquet operator 
${\cal V}$ where $M = {\rm Int}[\hbar_{\rm eff}^{-1}]$. A particle initially
in a state $\psi_0$ occupies a state $\psi_\tau = {\cal V}^\tau \psi_0$ 
after $\tau$ collisions at the cavity's boundary. The transport properties of 
this system derive from its scattering matrix \cite{markus}
\begin{eqnarray}\label{blocks}
{\cal S}= \left( \begin{array}{ll}
{\bf r} & {\bf t}' \\
{\bf t} & {\bf r}' 
\end{array}\right),
\end{eqnarray}
which we write in terms of $N \times N$ transmission (${\bf t}$) and 
reflection (${\bf r}$) matrices. From ${\cal S}$, the system's conductance is 
given by $g={\rm Tr} ({\bf t}^\dagger {\bf t})$. To construct ${\cal S}$ 
from ${\cal V}$, we couple the cavity to the left (L) and right (R) leads by 
introducing a $2N \times M$ projection matrix $P=P^{\rm (L)}+P^{\rm (R)}$.
Expressed in the basis of channel modes, the projection matrices read 
$P^{\rm (L,R)}_{nm}=1$ if $m=n \in \{m_i^{\rm (L,R)} \}$ and 
$P^{\rm (L,R)}_{nm}=0$ otherwise. The sets 
$\{m_i^{\rm (L,R)}\}$ are the $N$ components of ${\cal V}$ ideally connected 
to the modes of the L or R lead respectively. The energy-dependent ${\cal S}$ 
becomes \cite{fyodorov}
\begin{subequations}\label{smatrix}
\begin{eqnarray}
{\cal S}(\varepsilon) 
&=& \sum_{\tau=0}^\infty \exp[i (\tau+1) \varepsilon] {\cal S}_\tau,\\
{\cal S}_\tau(\varepsilon) &=& P
[{\cal V} (1-P^T P)]^{\tau} {\cal V}P^T.\\[-4mm]
\nonumber
\end{eqnarray}
\end{subequations}
The $\tau^{\rm th}$ term in the above Taylor expansion corresponds to the 
time-evolution operator for a particle colliding exactly 
$\tau$ times at the boundary of the cavity before exiting. 

\begin{figure}
\centerline{\hbox{\includegraphics[width=0.85\columnwidth]{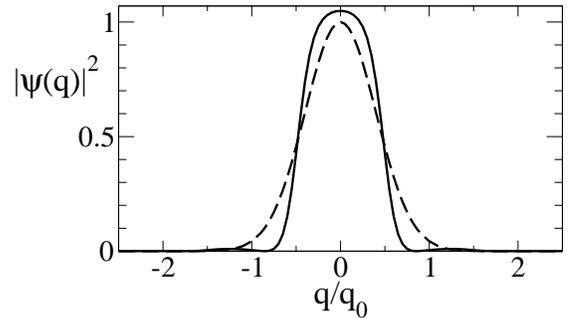}}}
\caption[]{\label{figure1}
Squared amplitude of a PS state (solid line)
versus the normalized coordinate $q/q_0$, $q_0= L\sqrt{2 \pi \hbar_{\rm eff}}$.
For comparison we also show a coherent state (dashed line).}
\vskip -0.1truein
\end{figure}

To maximally resolve the underlying classical dynamics, we wish to 
express ${\cal S}$ in an appropriate orthonormal basis. We start from 
a complete set of coherent states centered on a discrete
von Neumann lattice \cite{perelomov}. While this set is not orthogonal,
it is nevertheless complete (and not overcomplete). 
It is then possible to recursively construct an orthonormal Phase Space (PS) 
basis from it, $\{|(p,q)_i\rangle_{\rm L,R} \}$,
where each PS basis state is exponentially localized in both 
position and momentum and covers a phase-space area ${\cal O}(\hbar)$
\cite{caveat0a}.
The algorithm for constructing this basis is similar to those used in wavelet 
analysis \cite{wavelet-book}, and we will discuss it elsewhere
\cite{jacquod}. Here we only show such a PS state in
Fig.~\ref{figure1}. The transformation from lead modes to this orthonormal 
basis is a unitary transformation 
${\cal S} \rightarrow  {\cal U}^\dagger {\cal S} {\cal U}$
and hence leaves the transport properties of ${\cal S}$ unaffectedx.
We will now show that the multi-bounce expansion of Eqs.~(\ref{smatrix}), 
expressed in this orthonormal PS basis elegantly connects quantum transport 
to the underlying classical dynamics.

Two cases have to be considered when evaluating the matrix elements
$\langle (p,q)_i| {\cal S}_\tau |(p,q)_j \rangle$ of ${\cal S}_\tau$.
If $|(p,q)_j \rangle$ sits on classical trajectories
which all leave the cavity at the same time and through the same lead,
then its quantum mechanical time-evolution
is given by the classical one. This requires first that the classical 
trajectory starting at $(p,q)_j$ leaves the cavity after a time 
$\tau < \tau_{\rm E}^{\rm o}$ \cite{Zaslavsky}
(this ensures that ${\cal S}_\tau |(p,q)_j \rangle$ is still a well
localized wavepacket), and second that at the time of escape 
the wavepacket is not located too close to the edge
of the lead, and so avoids partial reflection. When these two conditions are 
fulfilled, the whole wavepacket escapes the cavity 
at a single time $\tau$ through either the L or R lead (not both)
in a state ${\cal S}_\tau |(p,q)_j \rangle$. 
In all other instances this is not the case. For wavepackets that exit the 
cavity either at times $\tau > \tau_{\rm E}^{\rm o}$ or 
after one or more partial reflections at the edge of a lead,
diffraction and quantum coherence play an important role.
Thus we split ${\cal S}$ into a short- and a long-time
contribution \cite{caveat0},
\begin{subequations}\label{psmelsplit}
\begin{eqnarray}
\langle (p,q)_i| {\cal S}|(p,q)_j \rangle & = &
\langle (p,q)_i| {\cal S}^{\rm cl}+{\cal S}^{\rm qm}|(p,q)_j \rangle, \\
{\cal S}^{\rm cl}=\sum_{\tau < \tau_{\rm E}^{\rm o}}
{\cal S}_\tau  & ; &
{\cal S}^{\rm qm} =
\sum_{\tau>\tau_{\rm E}^{\rm o}} {\cal S}_\tau.
\end{eqnarray}
\end{subequations} 

\begin{figure}
\centerline{\hbox{\includegraphics[width=0.8\columnwidth]{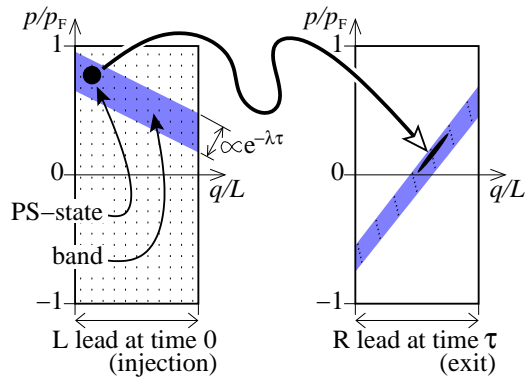}}}
\caption[]{\label{figure2} (color online)
Phase space of the injection (Left) and exit (Right) leads. The shaded area 
depicts a classical transmission band which goes from L to R in 
time $\tau$. Superimposed on the L lead is a square von Neumann lattice with 
spacing $(2\pi\hbar_{\rm eff})^{1/2}$. A PS state is placed at every vertex 
on the lattice, forming a complete orthonormal basis.
Each PS state initially covers a circle of radius ${\cal O}
[\hbar_{\rm eff}^{1/2}]$ in phase space. For $\tau < \tau_{\rm E}^{\rm o}$,
this state is still localized, but is stretched 
under the dynamics, as indicated by the dark ellipse on the R lead.}
\vskip -0.1truein
\end{figure}

The classical Liouville conservation of PS
volume guarantees that ${\cal S}^{\rm cl}$ and ${\cal S}^{\rm qm}$
act on different, nonoverlapping subsets of the PS basis.
To see this, we recall that the evolution of wavepackets 
that are initially localized in both position and momentum
is very well approximated semiclassically as a sum over classical 
trajectories \cite{steve,caveat4}. Thus if a classical
trajectory contributes to ${\cal S}^{\rm cl}$, then it does not
contribute to ${\cal S}^{\rm qm}$. From this {\it Liouville blocking},
we conclude that the PS states divide into two non-overlapping 
subsets (up to exponentially small corrections)
(i) the fully classical ones which leave the cavity after short 
times $\tau < \tau_{\rm E}^{\rm o}$ and are transmitted or reflected 
deterministically, and (ii) the quantum ones which
stay inside the cavity for longer times $\tau>\tau_{\rm E}^{\rm o}$ and/or 
are partially reflected and transmitted at collision at the edge of a lead
\cite{caveat1}. From the dwell time distribution 
$\rho(\tau)=\tau_{\rm D}^{-1} \exp[-\tau/\tau_{\rm D}]$
in chaotic cavities, one gets a number 
$N_{\rm cl}=N (1-\exp[-\tau_{\rm E}^{\rm o}/\tau_{\rm D}])$ 
of states in subset (i) and
$N_{\rm qm}=N \exp[-\tau_{\rm E}^{\rm o}/\tau_{\rm D}] 
\propto \hbar_{\rm eff}^{-1+1/\lambda \tau_{\rm D}} \gg 1$ 
of states in subset (ii). These considerations lead to our first result that 
${\cal S}$ has the block-diagonal decomposition
\begin{eqnarray}\label{Ssplitting}
{\cal S} & = & {\cal S}^{\rm cl} \oplus {\cal S}^{\rm qm},
\end{eqnarray}
where ${\cal S}^{\rm cl,qm}$ both have the structure given in 
Eq.~(\ref{blocks}). Thus the transmission matrix ${\bf T}={\bf t}^\dagger
{\bf t}$ is also block-diagonal 
\begin{eqnarray}\label{Tsplitting}
{\bf T} & = & {\bf T}^{\rm cl} \oplus {\bf T}^{\rm qm},
\end{eqnarray}
with $N_{\rm cl,qm}$ by $N_{\rm cl,qm}$ matrices ${\bf T}^{\rm cl,qm}=
({\bf t}^{\rm cl,qm})^\dagger {\bf t}^{\rm cl,qm}$. Accordingly, the system's 
conductance is given by the sum of two contributions
\begin{eqnarray}\label{gsplitting}
g & = & g_{\rm cl} + g_{\rm qm},
\end{eqnarray}
where $g_{\rm cl} = N_{\rm cl}/2$ and $g_{\rm qm} \simeq N_{\rm qm}/2$.
Note that $g_{\rm qm}$ gives the contribution to the total
conductance carried by the stochastic scattering states.
As we will see below, it contains the weak localization corrections.
Eqs.~(\ref{Ssplitting}-\ref{gsplitting}) give a
microscopic justification of the two-phase fluid model
previously postulated in Refs.\cite{silvestrov,eugene}, where
the finite $\tau_{\rm E}^{\rm o,c}$ transport 
properties of a quantum chaotic system are given by
those of two separated cavities put in parallel. 
The decomposition (\ref{Ssplitting}) of ${\cal S}$ is naturally
connected to the underlying classical dynamics.
In fact, because of the finiteness of $\tau_{\rm D}$,
classical trajectories injected into a cavity
are naturally grouped into PS transmission and reflection 
bands \cite{silvestrov}, despite the ergodicity of the associated closed 
cavity. These bands are best visualized by considering
PS cross-sections of the left and right leads, and plotting
the initial and final coordinates of trajectories transmitted from left 
to right. This is sketched in Fig.~\ref{figure2}. All trajectories within 
one band follow neighboring paths and exit at the same time $\tau$ through 
the same lead. Because of the chaotic classical dynamics, bands with longer 
escape times $\tau$ are narrower, having a width (and hence a PS area) 
scaling like $\propto \exp[-\lambda \tau]$.
The Ehrenfest time is the time at which this area becomes smaller
than $\hbar_{\rm eff}$, consequently
for times longer than $\tau_{\rm E}^{\rm o}$ no band is big enough
for a PS state to fit within it \cite{silvestrov}. The meaning
of Eq.~(\ref{Ssplitting}) is now obvious:
The matrix ${\cal S}^{\rm cl}$ acts in the subspace of PS states
which are entirely inside a single band, while
${\cal S}^{\rm qm}$ acts in the subspace of PS states
which overlap many bands, exiting at different times, and
in general through different leads.

We now discuss the properties of ${\bf T}^{\rm cl}$ and ${\bf T}^{\rm qm}$.
The classical part of the transmission matrix
is $\sum_{\tau,\tau'<\tau_{\rm E}^{\rm o}} 
{\bf t}^{\dagger}_{\tau}{\bf t}_{\tau'}$. 
However each PS state in ${\bf t}^{\rm cl}$ follows a band that exits at a 
well defined time, thus only diagonal terms in the double sum over $\tau,\tau'$
are non-zero, and one has
\begin{eqnarray}\label{diag_tcl}
{\bf T}^{\rm cl}
&=&\sum_{\tau < \tau_{\rm E}^{\rm o}} 
{\bf t}^{\dagger}_{\tau} {\bf t}_{\tau}.
\end{eqnarray}
We stress that this is not the usual diagonal approximation 
and so we do not require any energy or ensemble averaging here. 
Once ${\bf T}^{\rm cl}$ is written as in Eq.~(\ref{diag_tcl}),
it is easy to show that it is diagonal.
To see this one recalls first that for $\tau < \tau_{\rm E}^{\rm o}$,
${\cal S}_\tau |(p,q)_j \rangle= |(\tilde{p},\tilde{q})_i \rangle$ 
is still a narrow wavepacket, which however has been stretched and
rotated by the classical dynamics \cite{Zaslavsky}
(this is sketched in Fig.~\ref{figure2}).
This can be included for each band individually 
into a $\tau$-dependent area-preserving coordinate transformation, so that
a basis made of these stretched PS states is complete and orthonormal within 
each band (relationships between bands can be ignored because Liouville 
blocking ensures that bands never overlap). This means that, after a 
$\tau$-dependent unitary transformation of the bra-states,
$\langle (\tilde{p},\tilde{q})_i| {\bf t}_{\tau}|(p,q)_j \rangle=0$ 
unless the state $|(p,q)_j \rangle$ in the L lead is the one state that
transmits to the (stretched and rotated) state 
$|(\tilde{p},\tilde{q})_i\rangle$ in the R lead, then
$\langle (\tilde{p},\tilde{q})_i| {\bf t}_{\tau}|(p,q)_j \rangle
=\exp[i \varphi_{ij}]$. Eq.~(\ref{diag_tcl}) shows that ${\bf T}^{\rm cl}$
is unaffected by this transformation, hence 
\begin{eqnarray}
{\bf T}^{\rm cl}_{ij}
\ =\ \sum_{\tau < \tau_{\rm E}^{\rm o}} \big| 
[{\bf t}_\tau]_{ij} \big|^2
\ =\  \delta_{ij} \times T_i, \;\; i \in \{1,\cdots N_{\rm cl}\},
\end{eqnarray}
with transmission eigenvalues 
$T_i$ which are either {\em zero} or {\em one}. While this result
was anticipated \cite{vanhouten}, 
to the best of our knowledge it is derived here for the first
time directly from a microscopic theory. Thus ${\bf T}^{\rm cl}$ cannot 
contribute to noise, moreover, all quantum phases (such as $\varphi_{ij}$) 
cancel in ${\bf T}^{\rm cl}$ and hence it carries no quantum coherence.
Both shot-noise and coherent effects like weak localization are carried solely 
by ${\bf T}^{\rm qm}$ which we now focus on.

Unlike in ${\bf T}^{\rm cl}$, the off-diagonal terms $\tau \ne \tau'$
strongly influence ${\bf T}^{\rm qm} =
\sum_{\tau,\tau'>\tau_{\rm E}^{\rm o}} 
{\bf t}^{\dagger}_{\tau}{\bf t}_{\tau'}$.
These terms give rise to coherence and to the 
pseudo-randomness of transmission phases, as many initial PS states
may now be partially transferred to the same final PS state. Conversely,
each initial PS state is partially transferred to many final PS states
\cite{Zaslavsky}. Taking the Liouville blocking into account, 
${\bf t}^{\rm qm}$ is ergodic within its own subspace
(being decoupled from ${\bf t}^{\rm cl}$). Because it
is made up of all the long-time contributions 
with $\tau>\tau_{\rm E}^{\rm o}$, it contains dynamically 
diffractive contributions necessary
for weak localization (Richter-Sieber pairs \cite{richter}). We further note
that its size $N_{\rm qm} \propto \hbar_{\rm eff}^{-1+1/\lambda \tau_{\rm D}}
\rightarrow \infty$ in the semiclassical limit.

This is all we need to generalize the semiclassical theory of weak 
localization to finite $\tau_{\rm E}^{\rm o,c}/\tau_{\rm D}$. 
The approach of Ref.~\cite{richter} can be followed
provided Eq.~(\ref{Ssplitting}) is taken into account by
considering an effective cavity with a number of modes $N_{\rm qm}$, 
a restricted phase-space area 
$\Sigma_{\rm qm}(E) = \Sigma(E) \exp[-\tau_{\rm E}^{\rm o}/\tau_{\rm D}]$ 
and a dwell time distribution 
$\rho_{\rm qm}(\tau) = \Theta (\tau-\tau_{\rm E}^{\rm o}) 
\exp[-(\tau-\tau_{\rm E}^{\rm o})/\tau_{\rm D}]/\tau_{\rm D}$.
A careful analysis of the dwell time for the 
Richter-Sieber pairs (and use of an optimal basis \cite{caveat0a})
yields $\de g= -1/4 (1 + {\cal O}[(\lambda\tau_{\rm D})^{-1}])$
independent of $\tau_{\rm E}^{\rm o,c}$ \cite{caveat3}.
Thus for good chaotic systems, $\lambda\tau_{\rm D}\gg 1$,
weak localization remains universal in the classical limit.

Turning our attention to shot-noise, we readily see that the
reduction of the fraction of quantum channels reproduces the predicted 
exponential suppression of the Fano factor in the large $\tau_{\rm E}^{\rm o}$
regime \cite{agam}, $F=\sum_n T_n(1-T_n)/\sum_n T_n \propto N_{\rm qm}/N
=\exp[-\tau_{\rm E}^{\rm o}/\tau_{\rm D}]$.
We expect there will soon be a semiclassical theory to explain the universal 
value $F=1/4$.
Indeed we believe that a theory of transmission along the lines of 
Ref.~\cite{haake} is possible, which would show
that the properties of ${\cal S}^{\rm qm}$ are captured by 
one of the circular ensembles of random matrices.
Assuming that this is the case, we have the full
distribution $P(T)$ of transmission eigenvalues 
$P(T) = \alpha P_{\rm rmt}(T) +(1-\alpha)
\left[\delta(T)+\delta(1-T)\right]/2 $, where
$P_{\rm rmt}(T) = \pi^{-1} [T(1-T)]^{-1/2}$ \cite{carlormt}
and $\alpha = \exp[-\tau_{\rm E}^{\rm o}/\tau_{\rm D}]$, 
as postulated in Ref.~\cite{silvestrov} and numerically observed in 
Ref.~\cite{eugene}. If confirmed analytically, this would give the full
counting statistics of the current for any value of 
$\tau_{\rm E}^{\rm o}/\tau_{\rm D}$, including the correct Fano 
factor $F=(1/4) \; \exp[-\tau_{\rm E}^{\rm o}/\tau_{\rm D}]$ 
\cite{agam,caveat2}.

In summary, we have constructed a theory for 
transport in quantum chaotic ballistic systems in the
regime of finite $\tau_{\rm E}^{\rm o,c}/\tau_{\rm D}$. Our theory 
confirms the separation of the system into two subsystems, and thus
provides a microscopic foundation for the two-phase fluid model 
\cite{eugene,silvestrov}. Weak localization is predicted to remain universal 
in lowest order in $1/\lambda \tau_{\rm D}$, for any $\hbar_{\rm eff}$, in 
agreement with numerical data \cite{jakub4}, but in contradiction with earlier 
theories \cite{aleiner1,inanc}. Whether ${\cal S}^{\rm qm}$
is a random matrix remains an open question.

We are grateful to E. Sukhorukov for very interesting discussions and 
thank \.I. Adagideli, C. Beenakker, M. Sieber and P. Silverstrov for helpful 
comments. This work is supported by the Swiss National Science Foundation.

\bibliographystyle{apsrev}

\begin{thebibliography}{99}
\bibitem{caveat0a} This basis (chosen for pedagogical reasons) underestimates 
$N_{\rm cl}$. Both this basis and an optimal choice will be 
discussed at length in Ref.~\cite{jacquod}.
\bibitem{caveat0} In Eqs.~(\ref{psmelsplit}) it is understood that the 
$\tau < \tau_{\rm E}^{\rm o}$, part of ${\cal S}$ that leads to partial 
reflection at the edge of a lead is included in ${\cal S}^{\rm qm}$. 
This part has a subdominantly small size 
${\cal O}(\hbar_{\rm eff}^{-1/2}) \ll N_{\rm qm}$.
\bibitem{caveat4} This is true up to algebraically long times
$\tau^* = {\cal O}(\hbar_{\rm eff}^{-a}) \gg \tau_{\rm E}^{\rm o,c}$
($a > 0$, typically $a \approx 1/2$) \cite{steve}.
In the semiclassical limit $\hbar_{\rm eff} \rightarrow 0$,
this approach captures the transport properties arising from 
all but an exponentially small number of modes
$\propto \hbar_{\rm eff}^{-1}\exp[-1/\hbar_{\rm eff}^a \tau_{\rm D}]$.
\bibitem{caveat1} The basis transformation from channel modes to PS states 
can only be made unitary if the PS states which sit on the edge
of the leads are replaced by other states which are entirely
within the lead. These edge-of-lead states will be diffracted 
and so are included in ${\cal S}^{\rm qm}$ as in \cite{caveat0}.
This will be discussed at length in \cite{jacquod}.
\bibitem{caveat3} This result is easily generalized to asymmetric 
leads \cite{jacquod}.
\bibitem{caveat2} Contrary to what is commonly implied, the shot-noise power 
at fixed bias does not vanish in the limit
$\tau_{\rm E}^{\rm o}/\tau_{\rm D}\rightarrow \infty$ 
since it depends on $N_{\rm qm}$, and not on $N_{\rm qm}/N$.
\end{thebibliography}

\end{document}